\documentclass[floats,aps,prl,showpacs,preprint]{revtex4-1}

\usepackage{graphicx}

\begin{document}
\title{To wet or not to wet? Dispersion forces tip the balance for water-ice on metals}
\author{Javier Carrasco$^{1,2}$, Biswajit Santra$^{2}$, Ji\v{r}\'{i} Klime\v{s}$^{1}$,
Angelos Michaelides$^{1}$}
\affiliation{
$^1$London Centre for Nanotechnology and
Department of Chemistry, University College London, London WC1E 6BT, UK\\
$^2$Fritz-Haber-Institut der Max-Planck-Gesellschaft, Faradayweg 4-6, D-14195
Berlin, Germany\\
}

\begin{abstract}

Despite widespread discussion, the role of van der Waals dispersion forces in wetting 
remains unclear. 
Here we show that non-local correlations contribute substantially to the water-metal 
bond and that this is an important factor in governing the relative stabilities 
of wetting layers and 3D bulk ice.
Due to the greater polarizability of the substrate metal atoms, non-local correlations between water and the 
metal exceed those between water molecules within ice.
This sheds light on a long-standing problem, wherein common density functional theory 
exchange-correlation functionals incorrectly predict that none of the 
low temperature experimentally characterized ice-like wetting layers are thermodynamically stable. 

\end{abstract}


\maketitle


Water covers almost all solid surfaces under ambient conditions. 
As such, interfacial water is of crucial importance
to an endless list of problems in the physical and chemical
sciences. 
Examples include heterogenous ice nucleation on aerosol particles
of relevance to the atmospheric sciences, 
the flow of confined water through membranes and pores
in connection with geology and waste water treatment, or
the response and reactivity of water to electrochemical fields
\cite{hodgson09,henderson02}.
A prerequisite to understand these varied phenomena is 
the seemingly simple task of establishing how the water molecules are arranged 
at the interface, $i.e.$, what the water overlayer structure is.
Characterizing water overlayer structures is, however, a challenging task and 
despite thousands of publications on 
the chemical physics of water at interfaces only a handful of 
determinations have been accomplished to date
\cite{hodgson09,henderson02}.
%
These have all been on well-defined atomically smooth substrates -- mostly
metal surfaces -- under ultra-high vacuum (UHV) conditions,
to which the whole arsenal of surface science experimental probes
can be applied  \cite{hodgson09,henderson02}.
Instrumental in recent determinations has been density functional theory (DFT),
and indeed there is now an almost symbiotic relationship between 
DFT and scanning tunnelling microscopy
(STM), with DFT being used to provide structural models with which the experiments
can be interpreted.
Notable examples where DFT has been crucial in unravelling water 
overlayer structures include water in the submonolayer regime 
on Pd(111) \cite{cerda04}, Cu(110) \cite{carrasco09}, and Pt(111) \cite{nie10}.

Despite the undeniable value of DFT in helping to understand water on metals, 
there is an important and widely discussed problem \cite{feibelman02,cerda04,carrasco09,nie10,meng02,feibelman03,michaelides03,
michaelides04,meng04,meng05,ren06,
haq06,feibelman08,feibelman09,gallagher09,feibelman09b,hodgson09,schnur09,feibelman10,
feibelman10c}.
Specifically, the structures identified with widely used 
generalized gradient approximation (GGA) functionals such as PBE or PW91 are 
consistently found to be less stable than
ice Ih (the most common form of ice) \cite{feibelman02,cerda04,carrasco09,nie10,meng02,feibelman03,
michaelides03,michaelides04,meng04,meng05,ren06,haq06,feibelman08,feibelman09,gallagher09,schnur09,feibelman09b}.
This suggests that the intact water overlayers should simply not form
under equilibrium conditions.
This is an uneasy position for the role of DFT in water adsorption studies and it casts doubt on 
the structural characterizations themselves. 
%
%
Since the typical GGAs used
in such studies do not account properly for van der Waals (vdW) dispersion forces, it would 
of course be interesting and important to know what role vdW forces play in 
water adsorption. 
Apart from two limited attempts to address this issue \cite{feibelman05,hamada10}, 
and widespread discussion \cite{feibelman02,cerda04,carrasco09,nie10,meng02,feibelman03,
michaelides03,michaelides04,meng04,meng05,ren06,haq06,feibelman08,feibelman09,gallagher09,feibelman09b,hodgson09,schnur09,
feibelman10,feibelman10c},
no detailed study on the role of dispersion forces in 
realistic water adlayer structures on metals has been reported.
Thanks to the development of xc functionals which explicitly account for
dispersion, such as the non-local vdW density functional (vdW-DF) of Dion 
{\it et al.} \cite{dion04} and its offspring \cite{klimes10}, it is now possible to tackle this
question head on.

%
%

Here we report a study in which the role of dispersion in 
water metal bonding is examined in detail. 
To this end we apply a newly developed non-local functional to two of
the most widely studied water-ice adsorption systems, namely water on Cu(110) and 
Ru(0001). 
We find that non-local correlations contribute substantially to the water-metal 
bond and that this is an important factor in governing the relative stabilities 
of wetting layers and 3D bulk ice.
The fact that dispersion forces favor 1D and 2D wetting layer structures 
over 3D bulk ice is simply due to the much larger polarizability of 
the substrate metal atoms. 
This study highlights the key role dispersion plays in wetting
and how non-local functionals ameliorate the problem common GGAs have
with experimentally characterized wetting layers on metal surfaces.
Since quantitative first principles predictions of 
wetting and ice growth on metals are still beyond reach, some of the outstanding 
problems are also briefly discussed.
%


DFT calculations were performed with a modified version of the
VASP 5.2 code  \cite{vasp93,vasp96}, which includes our own 
self-consistent implementation
of the non-local van der Waals density functionals \cite{Soler}.
Results are reported for PBE \cite{pbe} and a modified version of vdW-DF, 
referred to as ``optB88-vdW'' \cite{klimes10}. 
The difference between the original vdw-DF of Dion $et$ $al.$ \cite{dion04} 
and optB88-vdW is merely in the exchange functional, 
with the optB88 exchange functional yielding more accurate interaction energies than
the original choice of revPBE \cite{klimes10}. 
The Cu(110) and Ru(0001) surfaces were modelled with 
slabs of between 4 and 6 layers thickness separated by 14 \AA\ of vacuum. 
Bulk ice was modelled using Hamann's twelve water molecule ice cell \cite{hamann97}.
Core electrons were treated with the projector augmented wave (PAW)
method \cite{paw}, whilst the valence electrons (3$d^{10}$4$s^1$ and  4$d^{7}$5$s^1$ 
for Cu and Ru, respectively) were expanded in a plane
wave basis with a 600 eV cut-off energy. 
Adsorption energies per water molecule were computed from
%
	$E_{\rm ads} = (E^{\rm tot}[{\rm H_2O/M}] - E^{\rm tot}[{\rm M}]- nE^{\rm tot}[{\rm H_2O}])/n$;
where $E^{\rm tot}[{\rm H_2O/M}]$ is the total energy
of the $n$ water molecule adsorption system, $E^{\rm tot}[{\rm M}]$ is the total energy of the
relaxed bare metal slab and $E^{\rm tot}[{\rm H_2O}]$ is the total energy of an 
isolated gas phase H$_2$O molecule. 
Likewise the lattice energy of bulk ice is obtained by subtracting the total energy of
the $n$H$_2$O ice supercell
from the total energy of $n$ gas phase H$_2$O molecules. 
Zero point
corrected versions of these quantities were also computed from harmonic vibrational
frequencies.
%
Some of the energy differences between the structures considered here are very small. 
However, we have carried out extensive convergence tests 
(some of which are reported in Tables SI and SII of the supporting information) which give us confidence
that our chosen computational set-up is sufficiently close to convergence 
that none of the conclusions reached here would be altered if yet more accurate settings were used. 

Let us first consider the wetting of Cu(110). 
Below 140 K and for a range of
temperatures and coverages in the submonolayer regime, water forms 1D
chains which run perpendicular
to the Cu(110) ridges. 
These chains have been observed by a number of 
groups \cite{yamada06,lee08,carrasco09} and although caution
must in general be exercised in considering
how kinetics dictates water film morphology in low temperature UHV
experiments, all existing experimental evidence suggests that the 
1D chains are the thermodynamically stable phase for water on Cu(110) at 
low coverage.
The structure of the chains has recently been characterized and shown to be
a periodic arrangement of water pentagons on the basis of STM, infra-red
spectroscopy, and DFT \cite{carrasco09}.
Hence the 1D chains provide an excellent opportunity to 
understand the relative stability of bulk ice with a well-defined wetting 
layer and to explore the role of dispersion forces in 
wetting on metals.

\begin{figure}[t!]
\includegraphics[width=0.95 \linewidth]{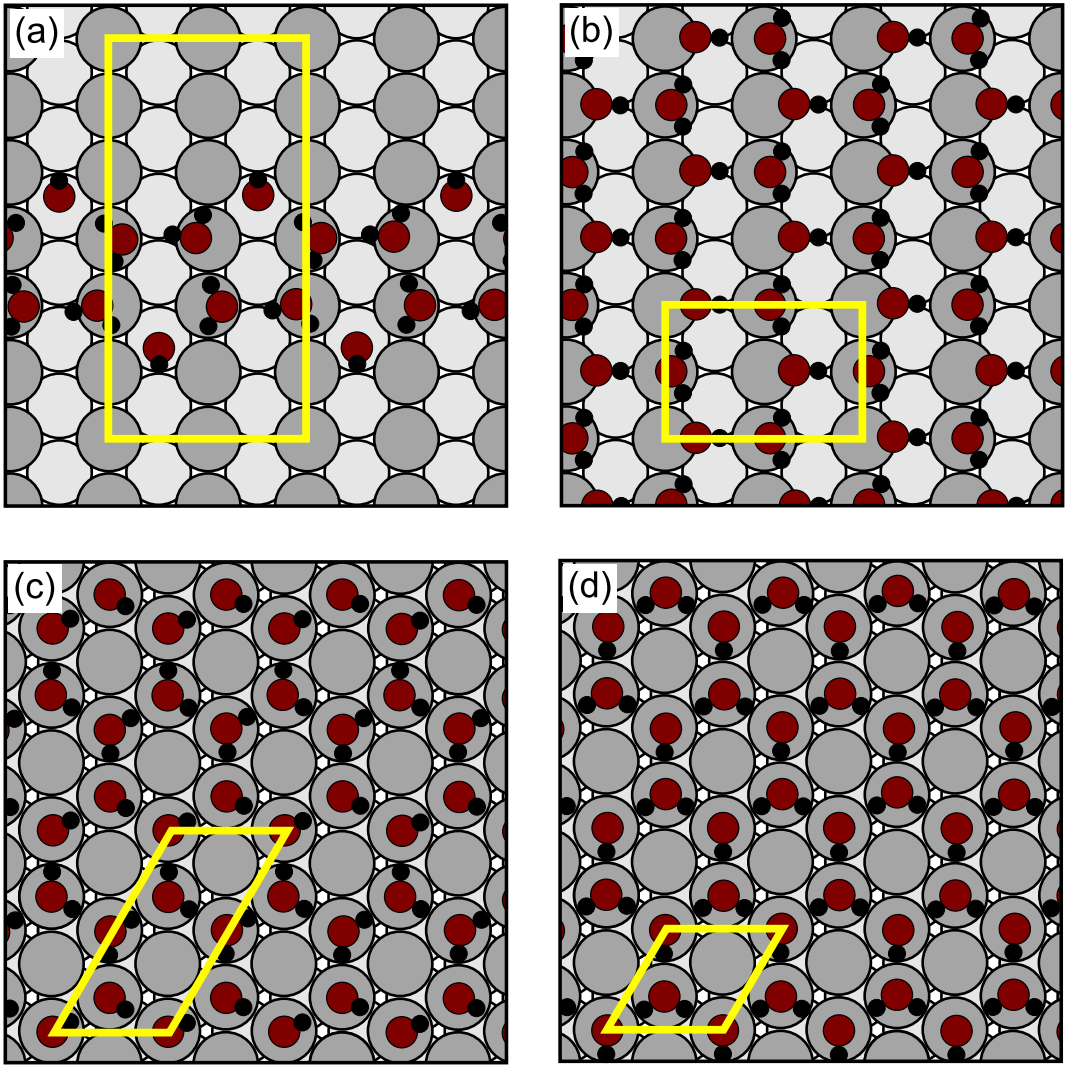}
\caption[]{Top views of water overlayers: (a) a pentagonal chain on Cu(110)  \cite{carrasco09};
(b) H-down bilayer on Cu(110); (c) extended chains on Ru(0001) with flat-lying and H-down
molecules  \cite{haq06}; 
and (d) H-down bilayer on Ru(0001).
 Black, red, dark gray, and light gray circles represent H, O, and Cu (Ru) in the first and second layer,
respectively. For molecules in the H-down configuration
the H atoms beneath the oxygens are not visible. The unit cells used are indicated.
\label{fig:models}}
\end{figure}

\begin{figure}[t!]
\includegraphics[width=0.95 \linewidth]{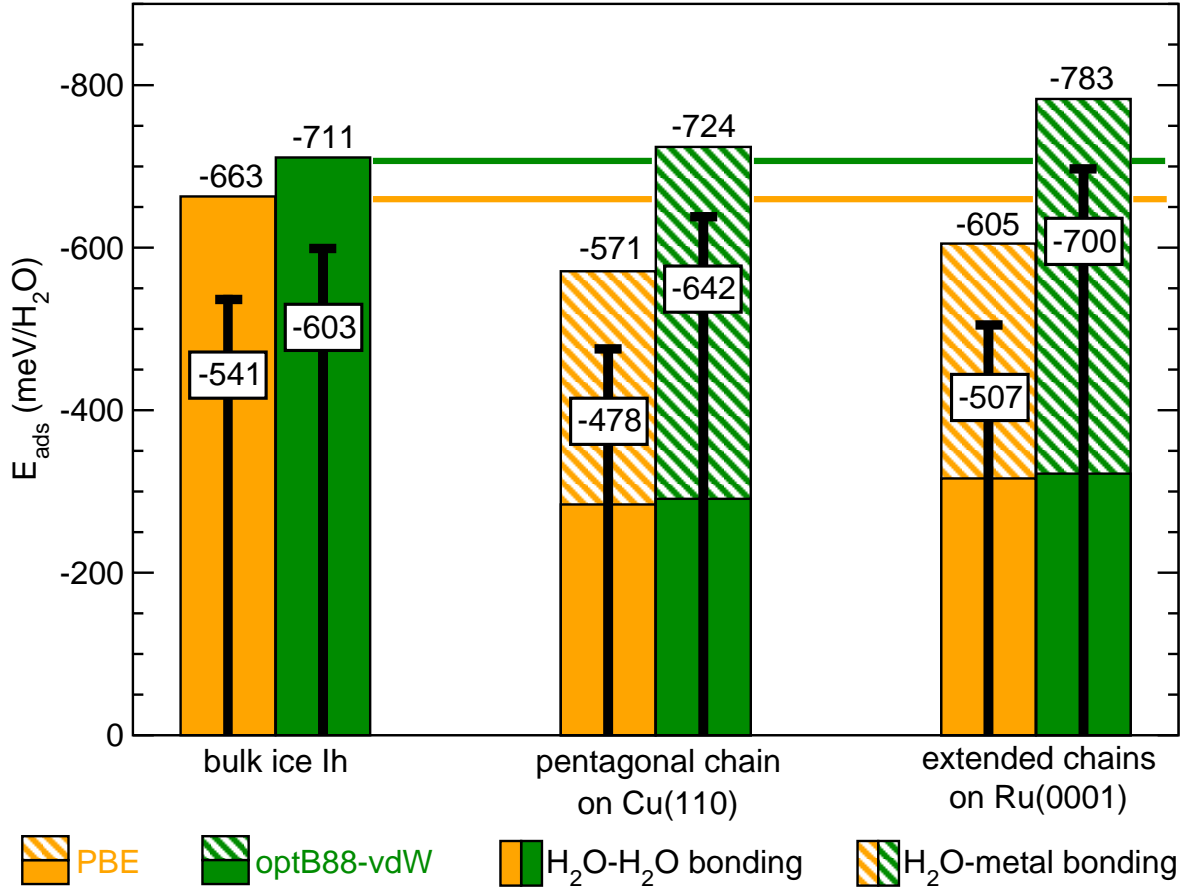}
\caption[]{Adsorption energies 
of water overlayers on Cu(110) and Ru(0001) compared to the
lattice energy of bulk ice Ih using PBE and optB88-vdW. 
The
adsorption energy is also decomposed in to water-water 
and water-metal bonding.
The thin centered black bars and associated numbers are the adsorption/lattice 
energies corrected for zero point energies. 
\label{fig:bar_plot}}
\end{figure}

The structure of the water chains is shown in Fig. \ref{fig:models}(a). 
They are comprised of a
face sharing arrangement of water pentagons with two types of water molecules:
flat-lying molecules bonded directly to Cu surface atoms along the 
ridges and upright molecules over the troughs in the (110) surface
which interact relatively weakly with the substrate.
Fig. \ref{fig:bar_plot} summarizes the adsorption energies obtained, along with the corresponding 
lattice energies of bulk ice. 
Results with and without zero point energies are 
reported. 
The first observation is that PBE predicts bulk ice to be more stable than the 
pentagon-based wetting layer. 
Specifically, the energy
of the adsorbed structure is 92 meV/H$_2$O less stable than the lattice
energy of bulk ice. 
Accounting for the zero point energy difference between the overlayer and ice is not sufficient to
correct the ordering of the two states,
although it does push things in the right direction by reducing the preference for bulk ice to 63 meV/H$_2$O.
This incorrect preference for ice is precisely the problem
that has been observed many times before for pure water overlayers computed 
with PBE and PW91 \cite{feibelman02,cerda04,carrasco09,nie10,meng02,feibelman03,
michaelides03,michaelides04,meng04,meng05,ren06,haq06,feibelman08,feibelman09,gallagher09,feibelman09b,hodgson09,schnur09,
feibelman10,feibelman10c}.
Moving to the optB88-vdW functional, we find that it does indeed give the correct energy ordering, 
predicting that the overlayer is marginally (13 meV/H$_2$O) more stable than bulk ice at the total energy
level and 39 meV/H$_2$O more stable when zero point energies are accounted for. 
Thus a stable wetting layer is found in agreement with the experimental observations. 
The structure of the overlayers obtained with PBE and
optB88-vdW are very similar (all hydrogen bonds between water molecules and water-metal
distances are within  0.02 \AA), which
indicates that whilst dispersion 
is important for the absolute adsorption energy of water it is of less
importance to the atomic structure.

As a consistency check, we compared the adsorption energy of the pentagonal overlayer
with extended 2D bilayers (both the ``H-down'', Fig. \ref{fig:models}(b), and ``H-up'' bilayers considered for 
this system before \cite{ren06}) and alternative 1D chain models \cite{carrasco09}.
Irrespective of the functional used the 
pentagonal chain structure remains the most stable overlayer. 
Therefore, the relative energies of the various overlayers are not 
strongly affected by dispersion forces, which
explains why PBE and PW91 have had success in predicting water structures on metals.
Finally, explicit consideration of finite temperature vibrational free energy
effects does not alter the conclusions drawn here and when these effects are accounted for the pentagons 
remain more stable than bulk ice (Fig. SII).


Regarding now the wetting of Ru(0001), Hodgson and co-workers have shown that between
140 and 160 K a 2D wetting layer of 
intact water molecules forms, which upon heating or 
exposure to electrons transforms in to a mixed H$_2$O-OH overlayer  \cite{haq06,gallagher09}.
At 0.67 monolayers the intact water overlayer exhibits 
a sharp $\sqrt{3}$ diffraction pattern in low-energy electron
diffraction (LEED) but a very low specular reflectivity in helium atom scattering (HAS).
Although the precise structure of the overlayer has not
been resolved, a compelling model that explains both the order observed in LEED 
and the apparent disorder from HAS  has been put forward. 
The overlayer involves water adsorbed near Ru atop sites in a hydrogen bonded honeycomb network, containing chains of
flat-lying and chains of H-down bonded water in a hexagonal superstructure.
%
%
%
Here, we have considered this extended chains model in
a  2$\sqrt{3}\times\sqrt{3}$ unit cell (Fig. \ref{fig:models}(c)), as well as 
intact H-down (Fig. \ref{fig:models}(d)) and H-up $\sqrt{3}\times\sqrt{3}$ bilayers.
With PBE the chains are more stable than the bilayers but
are less stable than bulk ice (Fig. 2 and Table SII) \cite{hodgson_comment}.
Therefore, as with water on Cu(110), with PBE one would not expect the formation of an  
extended overlayer.
However, when vdW interactions are accounted for the situation is
reversed. In particular, optB88-vdW predicts that the extended chain
structure
is indeed 72 meV/H$_2$O (97 meV/H$_2$O if zero point energy is
included) more stable than bulk ice. 
These results are in apparent agreement
with experiment, where wetting is observed for an intact water overlayer.
Again the structures obtained from PBE and optB88-vdW
do not differ to any great extent (all distances are within 0.02 \AA).

\begin{table*}[t!]
\centering
\caption{\label{tab:1} Different contributions to the optB88-vdW adsorption energy ($E_{\rm ads}$)
of the pentagons on Cu(110) and the extended chain structure on Ru(0001): water-water
bonding ($E_{\rm gas}^{\rm H_2O-H_2O}$),
water-metal bonding ($E_{\rm ads}^{\rm H_2O-M}$),
non-local correlation contribution to the adsorption energy
($E_{\rm ads}^{\rm nlc}$), non-local correlation contribution to water-water bonding 
($E_{\rm gas}^{\rm H_2O-H_2O,nlc}$), and non-local correlation contribution to water-metal
bonding ($E_{\rm ads}^{\rm H_2O-M,nlc}$). Zero point energy corrected values are given in parenthesis. All values in meV/H$_2$O.
}
\begin{ruledtabular}
\begin{tabular}{ccccccc}

                &    $E_{\rm ads}$ &  $E_{\rm gas}^{\rm H_2O-H_2O}$ & $E_{\rm ads}^{\rm H_2O-M}$ & $E_{\rm ads}^{\rm nlc}$  & $E_{\rm gas}^{\rm H_2O-H_2O,nlc}$ & $E_{\rm ads}^{\rm H_2O-M,nlc}$ \\
 \hline

Cu(110)    & --724 (--642)       & --291 & --433 & --488 & --128 & --360 \\                     
Ru(0001) & --783 (--700)       & --322 & --462 & --536 & --158 & --378 \\              

\end{tabular}
\end{ruledtabular}
\end{table*}

We now discuss why dispersion improves the relative energies of the overlayers and bulk ice. 
A decomposition of the total adsorption energy in to 
water-water and water-metal contributions proves useful (see the supporting information for details). 
With PBE the total adsorption energy is split equally
between water-water and water-metal bonding (Fig. \ref{fig:bar_plot}). 
When we switch to optB88-vdW, the adsorption energy
increases by about 150 and 180 meV for Cu and Ru, respectively. 
Crucially this 
increase comes almost exclusively from water-metal bonding,
with essentially no change in water-water bonding (Fig. 2). 
Further analysis shows that this increase in water-metal bonding is largely due to
non-local correlation. 
Indeed, on both Cu and Ru the non-local
contribution to the water-metal bonding exceeds that for the water-water bonding by 
a factor of 2 to 3 (Table \ref{tab:1}).
It can also be seen from Table \ref{tab:1} that the non-local contribution
to the water-metal bonding on Ru is larger than that on Cu, which explains why 
dispersion has a larger impact on Ru than on Cu. 
Although, it is somewhat counterintuitive that dispersion favours the formation
of 1D and 2D structures over 3D bulk ice, our analysis is consistent with
the fact that Ru and Cu atoms have much larger polarizabilities  than O and H. 
Specifically, the relative $C_6$ coefficients from time-dependent DFT calculations
\cite{grimme10} of Ru : Cu : O : H are 79 : 44 : 2 : 1, respectively.

Before concluding, we stress that although progress has been made we are still some way from
quantitative first principles predictions of wetting and related phenomena 
such as heterogeneous ice nucleation rates.
In particular, here we have obtained an improved description of the relative energies of 
water-ice overlayers and bulk ice, but this is achieved at the expense of 
a slightly worse absolute lattice energy for ice.
The experimental lattice energy for bulk ice Ih is 610 meV/H$_2$O \cite{feibelman08}, 
our PBE value is --663 meV/H$_2$O, and
our optB88-vdW value is --711 meV/H$_2$O.
PBE outperforms optB88-vdW on the lattice energy because the missing dispersion
is compensated for by the tendency of PBE to slightly overestimate H-bonds between
water molecules which are almost linear \cite{santra07}.
optB88-vdW, on the other hand, is known to overestimate the vdW correlation as the number of H-bonds between
water molecules increases \cite{klimes10}. 
Associated with this overbinding is a lattice constant that is too small: the experimental volume
is 32.1 \AA$^3$/H$_2$O, the PBE value is  30.4 \AA$^3$/H$_2$O, and the optB88-vdW value is  29.8 \AA$^3$/H$_2$O.
The issue of the lattice constant of bulk ice is particularly relevant to heterogeneous ice nucleation, where epitaxial
match between the substrate and ice has long been considered a key factor in the nucleating 
ability of a substrate. 
It remains to be seen if non-local vdW functionals in their current form are capable of simultaneously addressing
all of these issues. 
Analysis of the underlying errors in the exchange and (non-local) correlation
components might help to shed light on this \cite{Romaner}. 
Although, in the longer term the application 
of the random phase approximation within the adiabatic connection fluctuation dissipation theorem or 
quantum Monte Carlo to ice-like overlayers on metals would be welcomed.

In conclusion, we have considered the role of dispersion in water-metal bonding.
Analysis of the relative contributions of water-water 
and water-metal bonding, to two wetting layers of contemporary importance, highlights the critical role 
dispersion plays in the wetting of metals by water. 
Given that DFT plays a central role in interpreting and understanding 
the most well-defined experimental studies of wetting, 
it is satisfying that progress with the long-standing wetting problem has been made.
Although dispersion is of minor importance for many questions of structure and bonding, we have shown here 
that the dispersion forces
between water and metals are sufficiently large to favor formation of 1D and 2D wetting structures over 3D bulk ice. 
This arises from the much larger polarizability of the metal atoms compared to oxygen and hydrogen, and so it is
likely that dispersion will be of importance to water on metals in general and not just the exemplar
systems considered here.

JC acknowledges financial support from the Royal Society through a
Newton International fellowship and support from Matthias Scheffler.
AM is supported by the EURYI scheme
(www.esf.org/euryi), the EPSRC, and the European Research Council. 
We are grateful for computer time to UCL
Research Computing and the UK's national high performance computing service 
HECToR (from which access was partly obtained via the UK's Material Chemistry
Consortium,  EP/F067496). 
We also thank Peter Feibelman and Alexandre Tkatchenko for valuable 
feedback on a draft of this article.


\begin{thebibliography}{99}
\bibitem{hodgson_comment} In \cite{haq06} a slightly larger value
(--658 meV/H$_2$O) was reported for the adsorption energy of the extended chain 
structure (using PW91, ultra-soft pseudopotentials,
and a 396 eV cut-off). Here, we
find that with our most tightly converged set-up at the PBE level the
chains have an adsorption energy of --605 meV/H$_2$O, about 50-60 meV/H$_2$O
less stable than ice with PBE.
\end{thebibliography}
\end{document}